# Molecular Configuration Around Single Vacancy in Solid $CO_2$ at $T = 0$ K Studied by Monte Carlo Simulation Technique

Koji Kobashi

Former Special Researcher, Shinko Research Co., Ltd., Japan
31-440, Takasu-cho 2-1, Nishinomiya 663-8141, Japan

Abstract

Energetically minimum configurations of the first- and second-nearest neighbor (NN) $CO_2$ molecules surrounding a vacancy, created by removing a single molecule from the *Pa3* structure, were calculated by use of the Monte Carlo simulation technique at zero temperature. It was found that among the 1st NN molecules, only six NN molecules, closest to the oxygen atoms of the removed $CO_2$ molecule, significantly changed the positions and orientations, while the other six molecules were not influenced by the creation of the vacancy. The configurations of the molecules no longer had the three-fold symmetry that the *Pa3* structure possessed. The influence of the vacancy on the 2nd NN molecules was relatively small.

Crystals of simple molecules such as $CO_2$, $N_2$, $O_2$, $F_2$, and $I_2$ were intensively studied in 1980s because they have rotational degree of freedom unlike rare gas molecules.[1-5] Infrared and Raman spectra of rotational oscillations, called librations, at ambient and high pressures as well as pressure-induced structural phase transitions were of main interest. Such research was being declined in 1990s as most subjects of interest had been investigated. Instead, computer simulations of pure and mixed fluids of such simple molecules began to be studied extensively, and it is so until today.[6] Recent dissemination of high speed and low price desktop computers, however, enables us to study the subjects that have no longer been pursued in 1980s. The purpose of the present paper was to identify energetically stable configurations (the center-of-mass (CM) positions and the angles $\theta$ and $\phi$) of the molecules surrounding a single vacancy in solid $CO_2$ at $T = 0$ K by use of the Monte Carlo (MC) simulation technique.[7,8] The main focus was the configurational changes in the 1st nearest neighbor (NN) and the 2nd NN molecules due to the creation of a vacancy.

The intermolecular potential model used in the present paper was the Kihara core potential model V of Ref. 4. It assumes a linear rod core with zero-diameter and 2.21-Å length in a $CO_2$ molecule, and the intermolecular potential $V_c(\rho)$ is expressed by:

$$V_c = U_0 \left[ 2\left(\frac{\rho_0}{\rho}\right)^9 - 3\left(\frac{\rho_0}{\rho}\right)^6 \right] + U_{EQQ}, \qquad (1)$$

where $\rho$ is the shortest distance between the cores associated with different $CO_2$ molecules, $U_0$ = 232 K (1 K = 1.38065×10$^{-23}$ J), and $\rho_0$ = 3.27 Å (1 Å = 0.1 nm). In addition, a point electric quadrupole moment $Q$ = -4.8 × 10$^{-26}$ esu.cm$^2$ is placed in the CM of the molecule, and $U_{EQQ}$ is the electric quadrupole-quadrupole (EQQ) interaction energy that depends on both the distance $R$ between the molecular centers and the molecular orientations. To facilitate the computation, the potential of Eq. (1) was applied only between molecules where $R \leq 10$ Å, and the first repulsive potential term in Eq. (1) was dropped between molecules where 10 Å < $R \leq 15$ Å. Interactions between molecules of $R$ > 15 Å were ignored. This approximation is justified because the pairwise repulsive energy, *i.e.*, the first term of Eq. (1), is 0.001 K at $\rho$ = 10 Å, and since there are roughly 100 molecules at $R \sim$ 10 Å, the decrease of the molecular energy[9] by the above approximation is ~ 0.1 K, which is only 10$^{-5}$ of the molecular energy and hence negligible.

The crystal structure of $CO_2$ at $T$ = 0 K and ambient pressure is known to be face-centered cubic (*fcc*) with four molecules per primitive unit cell, and the molecules are oriented in one of the four body diagonal directions. It belongs to a space group *Pa3*.[10] The experimental lattice constant is 5.5544 Å.[11,12] The present simulations started from this crystal structure: the crystal unit for simulation was a cubic box with each edge consisting of eight primitive unit cells, hence the edge length of the box was 44.298$_0$ Å, and the total number of molecules associated with the box was 2048. To make a crystal virtually infinite in size, the periodic boundary condition was used.[7] The calculated energy of the 2048 molecules was minimum when the lattice constant was 5.5372$_5$ Å, 0.31% smaller than the experimental value. The calculated lattice constant was also used for the simulation of a crystal with vacancy, in which case a $CO_2$ molecule oriented along [111] direction at the origin was removed. Thus, the vacancy was surrounded by 12 NN molecules and six 2nd NN molecules, and the total number of molecules associated with the cubic box was 2047. Because of the periodic boundary conditions, the vacancies form a simple cubic lattice structure with a lattice constant of 44.298$_0$ Å, which is considered to be far enough for the vacancies to influence with each other. Under these circumstances, the terms of "a vacancy" and "a single vacancy" are used in the present paper. The initial randomization of the molecular CM positions and orientations was set small not to destroy the initial structure (see below). The position of a vacancy was always set at the origin and immobile throughout the simulation. A parallel displacement of all molecules, corresponding to the $k$ = 0 phonon, was not observed. The parameters presented so far are summarized in Table 1. Finally, it should be mentioned that a simulation of a $CO_2$ crystal without vacancy was also undertaken in the same manner as in the case of the crystal with vacancy for the purpose of comparison.

The molecular positions and orientations were initially randomized within the range of -0.2 Å < $\Delta x, \Delta y, \Delta z$ < 0.2 Å and -20° < $\Delta\theta, \Delta\phi$ < 20°. These parameter set will be denoted as

{$\Delta r$, $\Delta \psi$} = {0.2 Å, 20°}, hereafter. These values were made smaller for 14 and 18 times for the crystals with and without vacancy, respectively, as the simulation proceeded, and the final parameter set was {0.001 Å, 0.1°} in both cases.[13] The data of the last 20 jobs were used for analysis. Since the temperature was zero in the present simulation, the new molecular configuration was accepted only if the molecular energy in the new configuration was less than that of the current configuration. A single computational job contained 10,000 rounds of calculations over all molecules, and the total number of the jobs was 104. Each job took less than 10 minutes. The number of accepted configurations in each job was ~5,000 at the beginning, but increased and saturated to ~20,000 when {0.001 Å, 0.1°}. The 20,000 accepted number of configurations means that for every trial for a new molecular configuration, only 0.1% was accepted on average. This is extremely small compared with 20 to 50% for MC simulations of molecular fluids, but it is commonly known that the acceptance ratio is smaller as the medium for simulation is denser.[7,8] The number of acceptance increased discontinuously when {$\Delta r$, $\Delta \psi$} was decreased. A curious behavior was that given the parameter set {$\Delta r$, $\Delta \psi$}, the accepted number of configurations was either unchanged or increased only by 10 to 20% even though the number of rounds in a job was set ten times more than the standard number of rounds, 10,000. Such a situation was unchanged even though the internal function for random number generation was replaced with other functions. A possible cause of this is described below, but the actual cause is still under investigation.

Shown in Fig. 1 are the average molecular energies (in *log* units) measured from the average molecular energies in the initial structures with (curve in blue) and without (curve in red) vacancy as a function of the number of acceptance. The energy increased to ~1,030 K by initial randomization, and finally decreased to 0.09 and 0.2 K for structures with and without vacancy, respectively. Most of the tiny peaks in the curves were due to the changes in {$\Delta r$, $\Delta \psi$}. It appears in Fig. 1 that there is no further decrease of energy in both cases even though the simulations were continued. This does not seem to be reasonable because for the crystal with vacancy, the ultimate crystal energy in simulation should be lower than the initial crystal energy because the configurational relaxation of the molecules around the vacancy reduces the crystal energy. For the crystal without vacancy, the molecular energy should be ultimately the same as that of the initial *Pa3* structure. The calculated results in Fig. 1 show no sign of this inference. To investigate the cause, the data of the crystal without vacancy was first examined closely. It was then found that in the last 20 jobs, the average deviations in the CM along the *a*, *b* and *c* axes were -0.0006, 0.0016, and -0.0014 Å, respectively, and the average deviations in angles $\theta$ and $\phi$ were virtually zero. Thus, it appeared as if the molecular configurations well converged to the initial *Pa3* structure. However, the standard deviations were still finite: the standard deviations in the CM positions along the *a*, *b* and *c* axes were 0.0022, 0.0021, and 0.0022 Å, respectively, and those in angles $\theta$ and $\phi$ were 0.08° and 0.11°, respectively. Furthermore, in the molecular configurations of the last job (the 104th job), for instance, there were molecules in the bulk with the absolute values of the positional and angular deviations from the *Pa3* structure being ~0.02 Å (~ 0.5% of the NN distance) and $\Delta\phi$ ~ 0.2°, respectively. The average molecular energy over the last 20 jobs was 0.4 K above that of the *Pa3* structure, and the standard deviation was 4.1 K. These values seemed to be unchanged even though the simulation was continued further because in each of the last 20 jobs, the number of acceptance was ~20,000 and there was no sign of change in the number of acceptance, while the number of acceptance should decrease and ultimately be zero if the molecular configurations move toward the convergence to the *Pa3* structure. This implies that there are many molecular configurations with similar energies in the vicinity of the *Pa3* structure, and the molecular configurations move around these local energy minima no matter how many calculations are done, and there was no chance for all the molecules to take exactly the *Pa3* structure at once. Such a situation is the same for the crystal with vacancy. In the following description on the configurations of the 1st and the 2nd NN molecules, therefore, it has to be kept in mind that the molecular configurations in the calculated crystal structure are away from the *Pa3* structure albeit the deviations are small.

Table 2 summarizes the calculated energies of the 1st and the 2nd NN molecules of the crystals with and without vacancy to compare with the counterparts of the *Pa3* crystals with and without vacancy. The numbers listed in the columns of "Simulation" are the differences between the molecular energies averaged over the last 20 jobs and the molecular energies listed in the columns of "*Pa3* crystal". For convenience, the molecules are labeled according to the molecular numbers used in the actual computations (see Fig. 2), and their original orientations in the *Pa3* crystal are also listed in the Table. Let us first see the results on the crystal without vacancy listed in the column of "No vacancy". The initial energy of each molecule, *i.e.*, the molecular energy in the *Pa3* structure, was -6206.7 K. The average energy of the 1st NN molecules by simulation was -0.3 K from the initial energy. Since the average molecular energy over all molecules in the *Pa3* crystal is 0.2 K, as seen in Fig. 1, the negative energy difference of -0.3 K is due to the insufficient number of averaging. For the 2nd NN molecules, the average molecular energy was 0.6 K.

For the crystal with vacancy, the energy of the 1st NN molecule around the vacancy in the *Pa3* crystal was -5814.8 K that is 391.9 K higher than the energy of the crystal without vacancy. The molecular energies calculated by simulation are listed in the column of "Single vacancy" below "Simulation". Significantly large energy differences of ~ -30 K are seen for molecules 2, 3, 4, 255, 1824 and 2018. As seen in Fig. 2, these molecules are present in the [111] direction of the removed $CO_2$ molecule at the origin. On the contrary, the energies of the molecules 31, 32, 226, 227, 1794, and 1976 are very close to the molecular energy of the *Pa3* crystal with a single vacancy. These molecules are located on the side of the [111] oriented missing $CO_2$ molecule at the origin. This indicates that only the first group of the 1st NN molecules was affected by the creation of the vacancy. The energies of the 2nd NN molecules were lower than that of the $Pa3$ crystal, and there was no significant difference in energy whether they are in the $a$, $b$, or $c$ direction from the vacancy.

Table 3 lists the CM positions along the $a$, $b$, and $c$ axes and orientations ($\theta$ and $\phi$) of the NN molecules in the similar manner as in Table 2. The column "$Pa3$ crystal" on the right-hand side lists the CM positions and the orientations of the 1st and the 2nd NN molecules in the $Pa3$ crystal. The column "Simulation" on the left-hand side includes two kinds of data obtained by simulation and averaged over the last 20 jobs: the column "Single vacancy" lists the data of the crystal with vacancy, while the column "No vacancy" lists the data of the crystal without vacancy. Note that the values in the column "Simulation" are the differences from the corresponding values in the column "$Pa3$ crystal".

For the 1st NN molecules of the crystal without vacancy, the absolute values of the positional deviations from the $Pa3$ crystal structure along the $a$, $b$, and $c$ axes were ≤ 0.015 Å, approximately ≤ 0.4% of the NN distance (3.915 Å). The standard deviations of the CM positions along the $a$, $b$, and $c$ axes were ≤ 0.01 Å (not listed in Table 3). The absolute values of the positional and the standard deviations were ≤ 0.012 Å and ≤ 0.01 Å for the 2nd NN molecules, respectively. The absolute values of the angular deviations of $\theta$ and $\phi$ were ≤ 0.2° and ≤ 0.4°, respectively. In the simulation, therefore, the CM and the angular deviations in the crystal without vacancy were fairly small.

The results of the crystal with vacancy were markedly different from those of the crystal without vacancy. First, all deviations in the CM coordinates of molecules 2, 3, and 4 were negative, and those of 255, 1824, and 2018 were positive, indicating that these molecules moved toward the origin, *i.e.*, the center of vacancy, with the absolute values being ≤ 0.05 Å, approximately ≤ 1.3% of the NN distance. The angular changes of these molecules were roughly one order of magnitude greater than those in the crystal without vacancy. These configurational changes were the cause of the energetic stability of the same molecules, as seen in Table 2. Molecule 2 rotated by 4° about the $c$ axis, as shown in Fig. 3a, and this is the case for molecule 2018 that is crystallographically equivalent to molecule 4 (see Fig. 2). For molecules 3, 4, 255, and 1824, both $\theta$ and $\phi$ changed by roughly 3° and 2°, respectively. The orientational changes of molecules 3 and 255, and molecules 4 and 1824 are schematically depicted in Fig. 3b and c, respectively. By contrast, the angular changes of other NN molecules, 31, 32, 226, 227, 1794, and 1796, were roughly one order of magnitude smaller than the first group of molecules, and this is consistent with the results in Table 2 that the deviations in energy of those molecules were as small as those of the molecules in the $Pa3$ crystal without vacancy. Regarding the 2nd NN molecules, there was no noticeable change in the CM positions as compared with the crystal without vacancy, but the orientational deviations were definitely greater.

To confirm the above results on the crystal with vacancy, an additional simulation was undertaken in which only the NN molecules surrounding a vacancy were movable and the configurations of all other molecules were fixed in the *Pa3* structure. It was found that the configurations of the NN molecules changed in the similar and almost quantitative manner. This result endorses that the results of the simulation, presented in Table 3, are reasonable.

The crystal structures seen along the <111> direction are depicted in Fig. 4, where Fig. 4a and 4b are the views of the *Pa3* structure and the crystal with a vacancy at the origin, respectively. Carbon and oxygen atoms are depicted in brown and red in color, respectively. The atomic diameters are van der Waals radii. An arrow in Fig. 4a indicates the molecule at the origin. It appears like a sphere as it is oriented in the [111] direction. On the other hand, an arrow in Fig. 4b indicates molecule 2, in which case the orientational deviation is $\Delta\phi \sim 4.3°$.

In summary, the changes in the CM positions and the orientations of the molecules around a vacancy in solid $CO_2$ were calculated using the MC simulation technique, and compared with the crystal structure without vacancy. The present simulation showed that significant configurational changes occur for the three molecules and the equivalent three other molecules in the NN molecules located in proximity to the oxygen atoms of the removed $CO_2$ molecule at the origin, and the three-fold symmetry was no longer maintained around the vacancy. In addition, those molecules coherently moved toward the vacancy, and hence the vacancy was slightly contracted in the <111> direction. It should be noted here that a rough estimate of the standard deviations of CM positions and librations of $CO_2$ molecules due to quantum mechanical zero-point motion were ~0.1 Å and ~3.6°, respectively. In the real crystal, therefore, molecular positions and orientations are considerably smeared by the quantum effect. Even so, the stable molecular configurations around a vacancy determined in classical mechanics are important as the molecules are in motion around the configurations.

Acknowledgement

It is acknowledged that GNU gfortran (https://gcc.gnu.org/fortran/) was used on Ubuntu (https://www.ubuntu.com) for the compiler and VESTA (http://jp-minerals.org/vesta/en/) for the drawing of the crystal structures in the present work.

Table 1. Parameters used for simulation.

| | | | |
|---|---|---:|---|
| Box size | This work | $44.298_0$ | Å |
| Lattice constant | This work | $5.5372_5$ | Å |
| | Exp. | 5.5544 | Å |
| Total number of molecules in the box | | 2048 | |
| NN distance | This work | 3.9154 | Å |
| 2nd NN distance | This work | $5.5372_5$ | Å |
| | Exp. | 5.5544 | Å |
| Potential parameters | $U_0$ | 232.0 | K |
| | $\rho_0$ | 3.27 | Å |
| | $l$ | 2.21 | Å |
| Interaction range | core-core | 10 | Å |
| | van der Waals | 15 | Å |
| | EQQ | 15 | Å |

Table 2. Molecular energies.

| Molecule | Label | Original orientation | Simulation (deviation from Pa3) | | Pa3 crystal | |
|---|---|---|---|---|---|---|
| | | | Single vacancy | No vacancy | Single vacancy | No vacancy |
| 1st NNs | 2 | $<\bar{1}11>$ | -31.5 | -5.1 | -5814.8 | -6206.7 |
| | 3 | $<1\bar{1}1>$ | -31.1 | 1.7 | | |
| | 4 | $<11\bar{1}>$ | -33.7 | 1.4 | | |
| | 31 | $<1\bar{1}1>$ | -0.1 | -0.6 | | |
| | 32 | $<11\bar{1}>$ | 0.7 | -4.6 | | |
| | 226 | $<\bar{1}11>$ | 2.1 | 0.5 | | |
| | 227 | $<1\bar{1}1>$ | 3.6 | 2.3 | | |
| | 255 | $<1\bar{1}1>$ | -34.2 | 0.3 | | |
| | 1794 | $<\bar{1}11>$ | 0.1 | -0.1 | | |
| | 1796 | $<11\bar{1}>$ | 0.5 | -4.6 | | |
| | 1824 | $<11\bar{1}>$ | -29.1 | -1.3 | | |
| | 2018 | $<\bar{1}11>$ | -33.1 | 6.0 | | |
| | Average | | -15.5 | -0.3 | | |
| 2nd NNs | 5 | $<111>$ | -20.6 | -3.7 | -6088.4 | -6206.7 |
| | 29 | | -17.4 | 2.5 | | |
| | 33 | | -24.5 | 2.9 | | |
| | 225 | | -21.1 | 2.6 | | |
| | 257 | | -21.1 | -1.0 | | |
| | 1793 | | -19.5 | 0.6 | | |
| | Average | | -20.7 | 0.6 | | |

Table 3. Configurations of 1st and 2nd NN molecules.

| Molecule | Label | Original orien-tation | Simulation | | | | | | | | | | Pa3 crystal | | | | |
|---|---|---|---|---|---|---|---|---|---|---|---|---|---|---|---|---|---|
| | | | Single vacancy | | | | | No vacancy | | | | | Single vacancy and no vacancy | | | | |
| | | | CM position (Å) | | | Angle (degree) | | CM position (Å) | | | Angle (degree) | | CM position (Å) | | | Angle (degree) | |
| | | | $a$ | $b$ | $c$ | $\theta$ | $\phi$ | $a$ | $b$ | $c$ | $\theta$ | $\phi$ | $a$ | $b$ | $c$ | $\theta$ | $\phi$ |
| 1st NNs | 2 | $<\bar{1}11>$ | -0.008 | -0.053 | -0.019 | 0.2 | 4.3 | -0.004 | 0.009 | -0.006 | -0.1 | 0.1 | 2.769 | 2.769 | 0.000 | 54.7 | 135.0 |
| | 3 | $<1\bar{1}1>$ | -0.024 | -0.029 | -0.035 | 2.9 | -2.2 | -0.012 | 0.013 | -0.009 | 0.0 | -0.4 | 0.000 | 2.769 | 2.769 | 54.7 | -45.0 |
| | 4 | $<11\bar{1}>$ | -0.035 | -0.040 | -0.005 | -2.8 | 2.2 | -0.011 | 0.011 | -0.005 | -0.1 | -0.1 | 2.769 | 0.000 | 2.769 | 54.7 | -135.0 |
| | 31 | $<1\bar{1}1>$ | 0.021 | -0.011 | 0.019 | 0.2 | 0.3 | 0.002 | 0.009 | -0.010 | 0.0 | 0.1 | 0.000 | 2.769 | -2.769 | 54.7 | -45.0 |
| | 32 | $<11\bar{1}>$ | -0.007 | -0.023 | 0.009 | 0.2 | 0.3 | 0.000 | 0.010 | -0.011 | 0.1 | 0.0 | 2.769 | 0.000 | -2.769 | 54.7 | -135.0 |
| | 226 | $<\bar{1}11>$ | 0.003 | 0.002 | 0.017 | -0.2 | -0.1 | -0.008 | 0.013 | -0.009 | 0.0 | 0.1 | 2.769 | -2.769 | 0.000 | 54.7 | 135.0 |
| | 227 | $<1\bar{1}1>$ | -0.009 | -0.009 | -0.009 | 0.0 | 0.3 | -0.011 | 0.012 | -0.010 | -0.2 | 0.1 | 0.000 | -2.769 | 2.769 | 54.7 | -45.0 |
| | 255 | $<1\bar{1}1>$ | 0.034 | 0.003 | 0.043 | 3.1 | -2.3 | -0.003 | 0.008 | -0.013 | -0.2 | -0.1 | 0.000 | -2.769 | -2.769 | 54.7 | -45.0 |
| | 1794 | $<\bar{1}11>$ | 0.012 | -0.023 | -0.011 | -0.4 | -0.2 | -0.004 | 0.008 | -0.010 | -0.1 | 0.3 | -2.769 | 2.769 | 0.000 | 54.7 | 135.0 |
| | 1796 | $<11\bar{1}>$ | 0.020 | 0.000 | 0.000 | 0.0 | 0.3 | -0.010 | 0.007 | -0.013 | 0.1 | 0.1 | -2.769 | 0.000 | 2.769 | 54.7 | -135.0 |
| | 1824 | $<11\bar{1}>$ | 0.050 | 0.019 | 0.016 | -2.9 | 2.2 | 0.000 | 0.005 | -0.013 | 0.0 | -0.1 | -2.769 | 0.000 | -2.769 | 54.7 | -135.0 |
| | 2018 | $<\bar{1}11>$ | 0.022 | 0.029 | 0.029 | 0.1 | 4.1 | -0.006 | 0.007 | -0.015 | -0.1 | -0.2 | -2.769 | -2.769 | 0.000 | 54.7 | 135.0 |
| 2nd NNs | 5 | <111> | 0.000 | -0.023 | 0.008 | 0.9 | 0.5 | -0.013 | 0.010 | -0.010 | -0.1 | 0.0 | 0.000 | 0.000 | 5.537 | 54.7 | 45.0 |
| | 29 | | 0.014 | -0.001 | 0.002 | 0.8 | 0.4 | 0.005 | 0.009 | -0.012 | -0.1 | 0.1 | 0.000 | 0.000 | -5.537 | 54.7 | 45.0 |
| | 33 | | 0.003 | -0.010 | 0.000 | -0.1 | -1.2 | 0.000 | 0.008 | -0.004 | -0.3 | 0.0 | 0.000 | 5.537 | 0.000 | 54.7 | 45.0 |
| | 225 | | 0.013 | -0.014 | 0.011 | -0.1 | -1.1 | -0.008 | 0.009 | -0.009 | 0.0 | 0.1 | 0.000 | -5.537 | 0.000 | 54.7 | 45.0 |
| | 257 | | 0.008 | -0.019 | -0.001 | -0.7 | 0.7 | -0.004 | 0.008 | -0.005 | -0.1 | 0.1 | 5.537 | 0.000 | 0.000 | 54.7 | 45.0 |
| | 1793 | | 0.007 | -0.003 | 0.008 | -0.8 | 0.7 | -0.006 | 0.002 | -0.012 | -0.1 | 0.1 | -5.537 | 0.000 | 0.000 | 54.7 | 45.0 |

**Figure captions**

Fig. 1. Energy behaviors of crystals with and without vacancy: crystal with vacancy (red curve), crystal without vacancy (blue curve).

Fig. 2. Nearest neighbor molecular positions and labels.

Fig. 3. Orientational changes of NN molecules (a) 2 and 2018, (b) 3 and 255, (c) 4 and 1824.

Fig. 4. Crystals (a) with and (b) without a vacancy seen from <111> direction. See text for the meaning of the arrows.

Fig. 1. Energy behaviors of crystals with and without vacancy: crystal with vacancy (red curve), crystal without vacancy (blue curve).

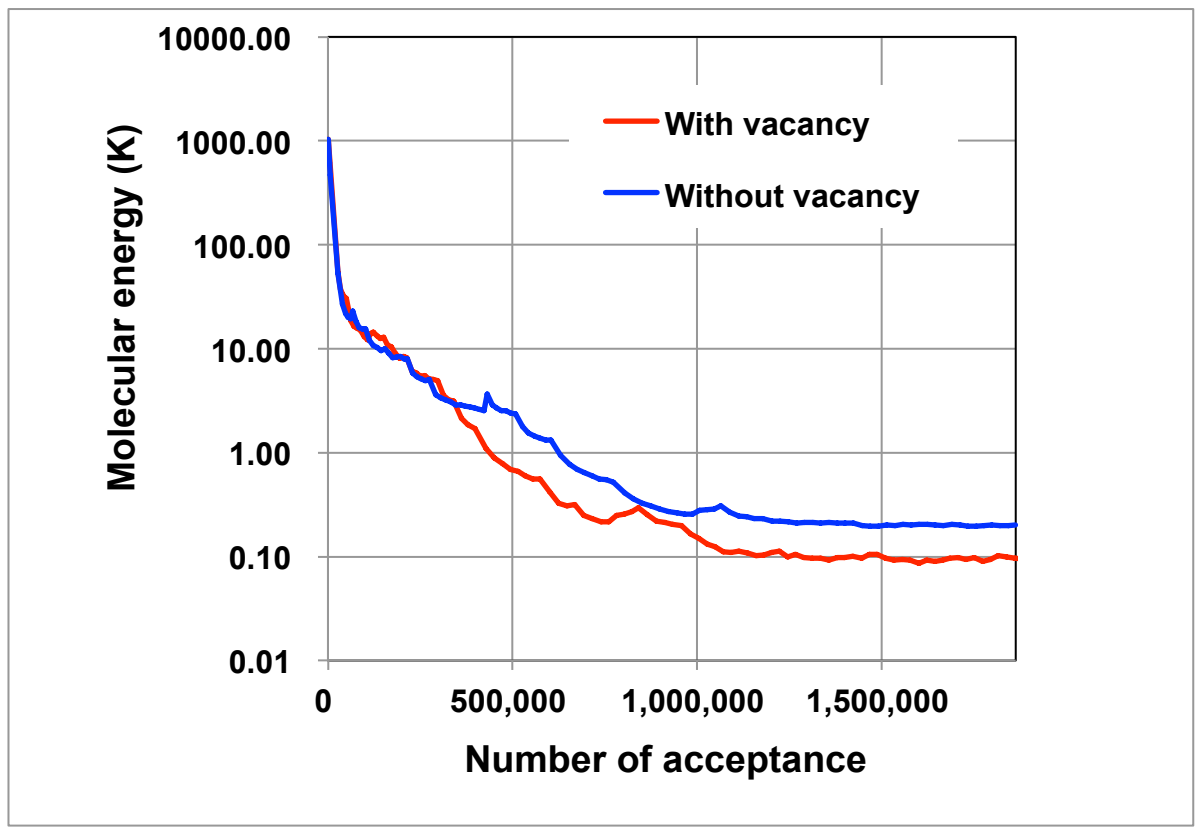

Fig. 2. Nearest neighbor molecular positions and labels.

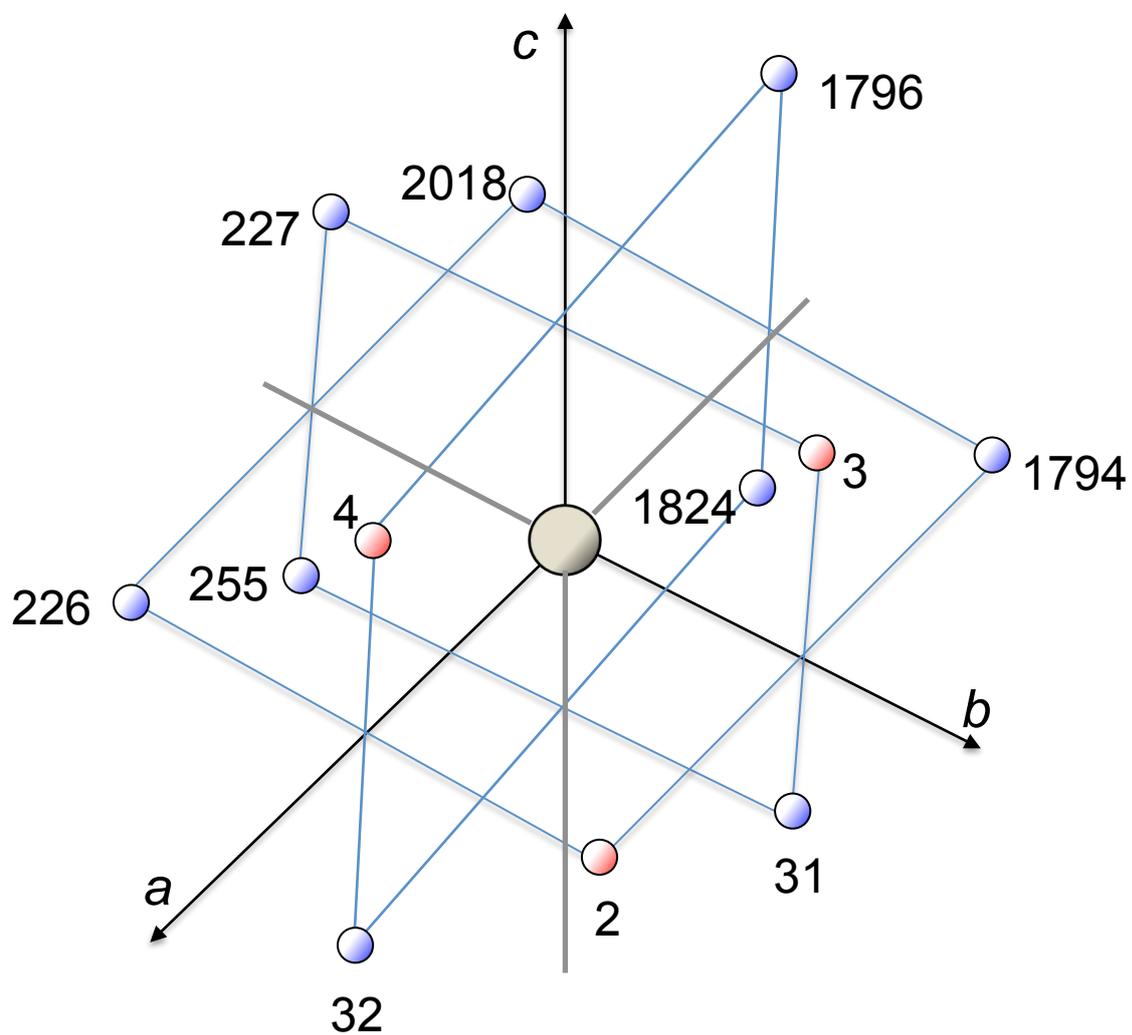

Fig. 3. Orientational changes of NN molecules (a) 2 and 2018, (b) 3 and 255, (c) 4 and 1824.

(a)

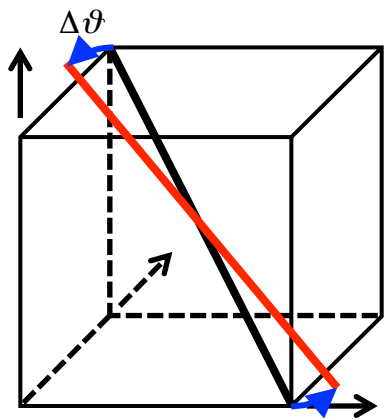

(b)

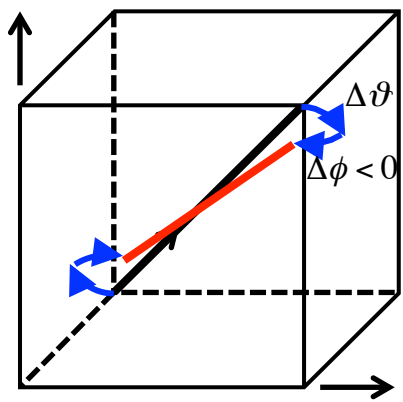

(c)

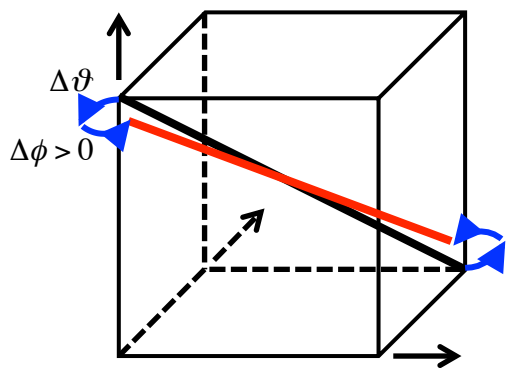

Fig. 4. Crystals (a) with and (b) without a vacancy seen from <111> direction.

(a)

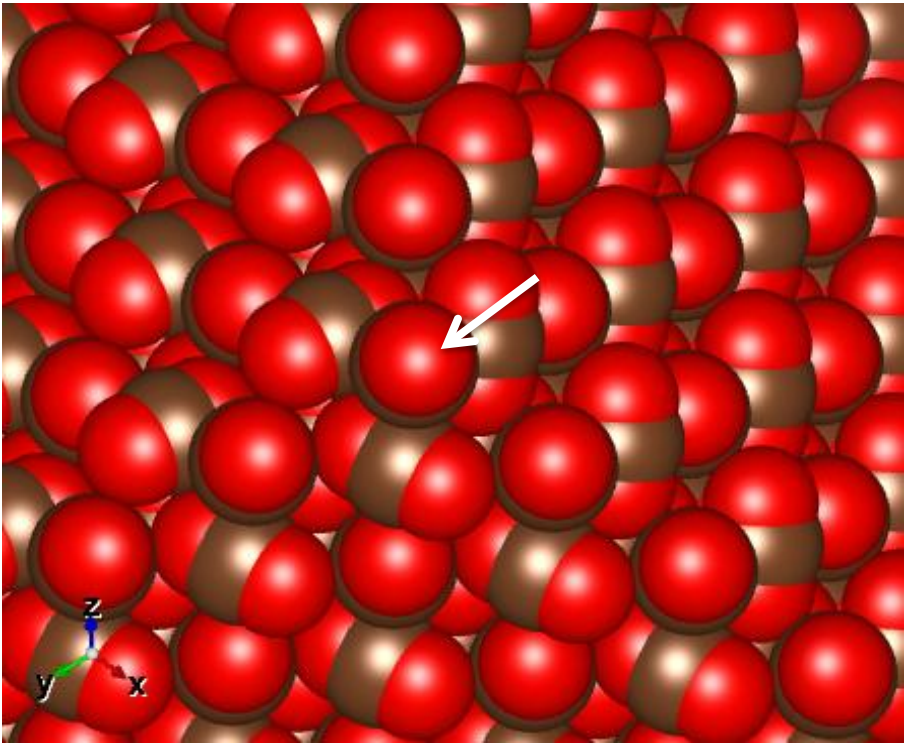

(b)

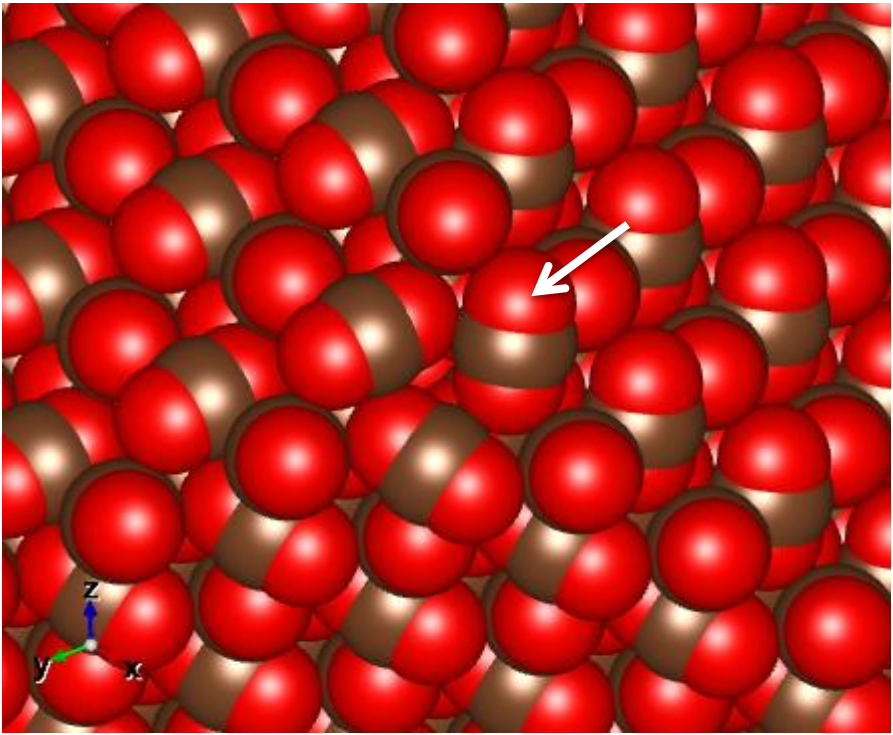